\documentclass[aps,twocolumn,showpacs, showkeys,nofootinbib,floatfix]{revtex4-1}

\usepackage{amsmath}
\usepackage{amsfonts}
\usepackage{txfonts}
\usepackage{graphicx}
\usepackage{dcolumn}
\usepackage{bbm}
\usepackage{amssymb}
\usepackage{latexsym}
\usepackage{titlesec}
\usepackage[colorlinks=true, linkcolor=red, citecolor=blue]{hyperref}
\usepackage{longtable}

\begin{document}

\title{Constraints on $f(R)$ Gravity through the Redshift Space Distortion}

\author{Lixin Xu$^{1,2}$}
\email{Corresponding author: lxxu@dlut.edu.cn}

\affiliation{$^{1}$Institute of Theoretical Physics, School of Physics \&
Optoelectronic Technology, Dalian University of Technology, Dalian,
116024, P. R. China}

\affiliation{$^{2}$State Key Laboratory of Theoretical Physics, Institute of Theoretical Physics, Chinese Academy of Sciences, Beijing, 100190, P. R. China}

\begin{abstract}
In this paper, a specific family of $f(R)$ models that can produce the $\Lambda$CDM background expansion history is constrained by using the currently available geometric and dynamic probes. The scale dependence of the growth rate $f(z,k)$ in this specific family of $f(R)$ model is shown. Therefore to eliminate the scale dependence of $f\sigma_8(z)$ in theory, which usually is defined as the product of $f(z,k)$ and $\sigma_8(z)$, we define $f\sigma_8(z)=d\sigma_8(z)/d\ln a$ which is obviously scale independent and reproduces the conventional definition in the standard $\Lambda$CDM cosmology. In doing so, under the assumption that future probes having the same best fit values as the current ten data points of $f\sigma_8(z)$, even having $20\%$ error bars enlarged, we find a preliminary constraint $f_{R0}=-2.58_{-0.58}^{+2.14}\times 10^{-6}$ in $1\sigma$ regions. 
This indicates the great potential that redshift space distortions have in constraining modified gravity theories. We also discuss the nonlinear matter power spectrum based on different halo fit models.
\end{abstract}


\maketitle

\section{Introduction}

In modern cosmology, it is common to realize the late time acceleration of our Universe either by considering an extra energy component namely \textit{dark energy} or modifying the theory of gravity. However, the two approaches are totally different in nature. Discrimination of dark energy from a modified gravity theory is a crucial issue in theory and cosmic probes. At the background level these are strongly degenerated as both yield the same expansion history of our Universe. But the dynamic evolution of a small perturbation would be different for different gravity theories. Therefore, observations of the large scale structure of our Universe may reveal some clue about the actual scenario. 

For the large scale structure, one can only read the correlation of galaxies, the tracers of the distribution of halos. And usually, one uses its Fourier transformation, the galaxy power spectrum $P_{g}(k)$. To understand the evolution of the matter perturbations even at the linear level, one should assume a relation between the overdensities of galaxy and matter, i.e., $\delta_g=b\delta_m$, where the overdensity for matter $\delta_m$ is well understood in theory. To compare the theory and cosmic observations, the so-called bias factor $b$, which usually depends on scales $k$, should be understood well. This galaxy bias issue limits the use of the matter power spectrum to study the large scale structure formation of Universe. For the nonlinear scale evolution, one still needs a better understanding of the halo model. The study of the nonlinear evolution through $N$-body simulation code with enough resolution and scales is numerically expensive and time-consuming. Thus, it is very difficult to scan model parameter space via the $N$-body simulation technique. In this regard, the so-called {\bf HALOFIT} model \cite{ref:Halofit} is an alternative plausible choice. 

The above argument is based on the observations to the continuity equation for the perturbation evolution. The other best thing is related to the velocity field which comes from the second perturbation equation, the so-called Euler equation. Although peculiar velocities are difficult to be observed directly, if galaxies can be treated as test particles, their peculiar velocities should be directly related to the total matter distribution. Actually, the galaxy maps will be distorted in the line of sight direction by peculiar velocities because of the interpretation of galaxy redshift as its distance. As a result, the overdensities in the redshift and real space are related via 
\begin{equation}
\delta^{s}_{g}({\bf k})=b\delta_{m}({\bf k})(1+\beta\mu^2),
\end{equation}
where $\beta=f/b$ is distortion factor, $f=d \ln \delta_{m}/d \ln a$ is the growth rate and $\mu=\cos(\theta_{kr})$, $\theta_{kr}$ being the angle between ${\bf k}$ and the line of sight. This is the redshift space distortion or the Kaiser effect \cite{ref:Kaisereffect}. Therefore, the combination $\sigma^{g}_{8}\beta=f\sigma_{8}$ is independent of galaxy bias in the linear case \cite{ref:Songfsigma8}. The redshift space distortion data are useful to constrain the cosmological parameters space \cite{ref:xufsigma8}. In this paper, we use the ten $f\sigma_8(z)$ data points as given in Table \ref{tab:fsigma8data} for constraining the model parameter space.
\begin{center}
\begin{table}[tbh]
\begin{tabular}{cccl}
\hline\hline 
$\sharp$ & z & $f\sigma_8(z)$ & Survey and Refs \\ \hline
$1$ & $0.067$ & $0.42\pm0.06$ & 6dFGRS~(2012) \cite{ref:fsigma85-Reid2012}\\
$2$ & $0.17$ & $0.51\pm0.06$ & 2dFGRS~(2004) \cite{ref:fsigma81-Percival2004}\\
$3$ & $0.22$ & $0.42\pm0.07$ & WiggleZ~(2011) \cite{ref:fsigma82-Blake2011}\\
$4$ & $0.25$ & $0.39\pm0.05$ & SDSS~LRG~(2011) \cite{ref:fsigma83-Samushia2012}\\
$5$ & $0.37$ & $0.43\pm0.04$ & SDSS~LRG~(2011) \cite{ref:fsigma83-Samushia2012}\\
$6$ & $0.41$ & $0.45\pm0.04$ & WiggleZ~(2011) \cite{ref:fsigma82-Blake2011}\\
$7$ & $0.57$ & $0.43\pm0.03$ & BOSS~CMASS~(2012) \cite{ref:fsigma84-Reid2012}\\
$8$ & $0.60$ & $0.43\pm0.04$ & WiggleZ~(2011) \cite{ref:fsigma82-Blake2011}\\
$9$ & $0.78$ & $0.38\pm0.04$ & WiggleZ~(2011) \cite{ref:fsigma82-Blake2011}\\
$10$ & $0.80$ & $0.47\pm0.08$ & VIPERS~(2013) \cite{ref:fsigma86-Torre2013}\\
\hline\hline
\end{tabular}
\caption{The data points of $f\sigma_8(z)$ measured from RSD with the survey references.}
\label{tab:fsigma8data}
\end{table}
\end{center}    

Because of the degeneracies between dark energy and a modified gravity model at the background level, we mainly focus on a specific family of $f(R)$ models, which produce the $\Lambda$CDM background expansion history \cite{ref:HeFR,ref:HeFRlinear,ref:HeFRnonliear}. Based on this model, the linear and nonlinear matter power spectrum were discussed in Refs. \cite{ref:HeFRlinear,ref:HeFRnonliear}, where the model parameter space was also constrained by the SDSS LRG matter power spectrum and the correlation between galaxy and the integrated Sachs-Wolfe effect (gISW). It was reported that CMB+SN+HST+MPK cannot constraint the model parameter space well, while tight constraints were obtained by taking gISW data into  account\cite{ref:HeFRlinear}. But these results were obtained at the risk of the galaxy bias issue, i.e., the understanding of the bias factor $b$ even at the nonlinear scale for the SDSS LRG data sets via the relation $P_{{\rm g}}(k)=(1+Qk^2)/(1+Ak)P_{{\rm lin}}(k)$, where $Q$ and $A$ are numbers needed to be calibrated. And it is crucial to calibrate these numbers $Q$ and $A$ for different cosmological models. For the gISW correlation data, we also have the bias parameter problem. However we should avoid the influence coming from a improper bias parameter $b$. The other risk comes from the nonlinear matter power spectrum, which is fitted by the {\bf HALOFIT} model based on $\Lambda$CDM model through $N$-body simulation. For a modified gravity model, this process should be repeated \cite{ref:HeFRnonliear,ref:MGhalofit}. Now it was already available for a range of model parameter $|f_{R0}| \lesssim10^{-4}$, named {\bf MGHalofit} \cite{ref:MGhalofit}, although it is based on Hu-Sawicki (HS) model \cite{ref:HSmodel}. It allows us to compare the nonlinear matter power spectrum between theory and cosmic probes, for instance considering the weak lensing, in a suitable range of model parameters. But the galaxy bias issue is still untouched.  

Considering above issues, in this paper, we try to use the well-understood linear perturbation to constrain a specific family of $f(R)$ models. One will see that the addition of RSD data sets can tightly constrain the model parameter space.   

We arrange this paper as follows. In Section \ref{sec:FRmodel}, we give a brief review of a specific family of $f(R)$ models. The constraint results will be shown in Section \ref{sec:results}, where we also give a discussion to the nonlinear matter power spectrum based on different halo fits. Section \ref{sec:conclusion} carries the concluding remarks.   

\section{A specific family of f(R) models} \label{sec:FRmodel}

\begin{center}
\begin{figure}[tbh]
\includegraphics[width=9.6cm]{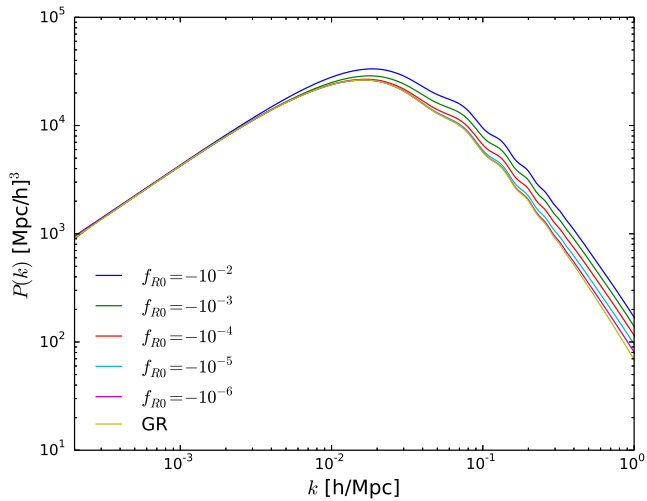}
\caption{The linear matter power spectrum at redshift $z=0$ for different values of $f_{R0}=-10^{-2}, -10^{-3}, -10^{-4}, -10^{-5}, 10^{-6}$ from the top to the bottom but the last, where the other relevant values are fixed to their mean values obtained in {\it Planck} 2013 \cite{ref:Planck2013CP}. The last one is for the $\Lambda$CDM model for comparison.}\label{fig:pks}
\end{figure}
\end{center}

The Einstein-Hilbert action in general form for $f(R)$ gravity reads as
\begin{equation}
S=\frac{1}{16\pi G}\int d^4x \sqrt{-g}\left[R+f(R)\right]+\int d^4x \sqrt{-g}\mathcal{L}_{m},
\end{equation}
where $\mathcal{L}_{m}$ is the Lagrangian of matter, which will not include the mysterious dark energy as  the late time accelerated expansion of our Universe can be realized by the proposed $f(R)$ gravity. For recent reviews for modified gravity theory, see \cite{ref:MGR1,ref:MGR2,ref:MGR3,ref:MGR4}. Doing variation with respect to the metric $g_{\mu\nu}$ for the Einstein-Hilbert action, one obtains a generalized Einstein equation which relates the geometry of space-time to the distribution of energy-momentum
\begin{equation}
F R_{\mu\nu}-\frac{1}{2}f g_{\mu\nu}-\nabla_{\mu}\nabla_{\nu}F+ g_{\mu\nu}\Box F =8\pi G T^{m}_{\mu\nu},
\end{equation}
where $F=1+\frac{\partial f}{\partial R}/\partial R$. It is obvious that the general relativity is recovered when $f(R)=0$ is chosen. The thorny problem is to determine a form of $f(R)$ which respects the cosmic observations. Here the the cosmic observations include two sides. One is the geometric, i.e., the expansion history of our Universe at the background level. The other is dynamic, i.e., the structure formation history of large scale of Universe via the linear and nonlinear perturbations. The $\Lambda$CDM model is compatible to almost all the cosmic observations at least at the background level. Therefore, an alternative cosmological model should not deviate from the $\Lambda$CDM model too much. This fact is also called cosmological model degeneracy. Thus, to discriminate one model from the other, reliable cosmic observations are demanded to break this degeneracy. The large scale structure formation information of Universe is promising to break the possible degeneracy because the structure formation history may differ significantly in models having same expansion history at the background level. Following this, one can detect a possible deviation from general relativity or rule out an alternative model. While constructing a model which predicts the expansion history as of $\Lambda$CDM model, one can compare the expansion rate and its time variation in the two models. A form of $f(R)$ having the expansion history as of $\Lambda$CDM model is \cite{ref:HeFR,ref:Nojiri} 
\begin{equation}
f(R)=-2\Lambda-\varpi\left(\frac{\Lambda}{R-4\Lambda}\right)^{p_{+}-1} {}_{2}F_{1}\left[q_{+},p_{+}-1;r_{+};-\frac{\Lambda}{R-4\Lambda}\right],
\end{equation}
where $\Lambda$ is the cosmological constant and $\varpi$ is a constant parameter
\begin{equation}
\varpi=\frac{D}{p_{+}-1}\left(\frac{\Omega_{m}}{\Omega_{\Lambda}}\right)^{p_{+}}3\Omega_{\Lambda}H^{2}_{0},
\end{equation}
and ${}_{2}F_{1}\left[a,b;c;z\right]$ is the Gaussian hypergeometric function and the indices are given by
\begin{equation}
q_{+}=\frac{1+\sqrt{73}}{12},\quad r_{+}=1+\frac{\sqrt{73}}{6},\quad p_{+}=\frac{5+\sqrt{73}}{12}.
\end{equation} 
This family of $f(R)$ models is specified by the only extra model parameter $D$ as a comparision to the $\Lambda$CDM model. This extra model parameter $D$ relates to the current value $B_0$ of the Compton wavelength 
\begin{equation}
B=\frac{f_{RR}}{1+f_{R}}\frac{d R}{d\ln a}\frac{H}{d H/d\ln a},
\end{equation}
via
\begin{eqnarray}
B_0&=&\frac{2D p_+}{(\Omega_m)^2\left\{1+D{_2F_1}\left[q_+,p_+;r_+;-\frac{\Omega_\Lambda}{\Omega_m}\right]\right\}}\nonumber\\
&\times&\left\{\frac{q_+}{r_+}\Omega_\Lambda{{}_2F_1}\left[q_++1,p_++1;r_++1;-\frac{\Omega_\Lambda}{\Omega_m}\right]\right.\nonumber\\
&-&\left.\Omega_m{{}_2F_1}\left[q_+,p_+;r_+;-\frac{\Omega_\Lambda}{\Omega_m}\right]\right\},\label{eq:b0}
\end{eqnarray} 
here $f_{RR}=\partial^2 f/\partial R^2$. It is also not difficult to find the relation between $D$ and $f_{R0}$ 
\begin{equation}
f_{R0}=D\times{{}_2F_1}\left[q_+,p_+;r_+;-\frac{\Omega_\Lambda}{\Omega_m}\right],\quad\label{mG}
\end{equation}
which is the current value of a new scale degree of freedom in this kind of modified gravity theory. Based on different values of $f_{R0}$, cosmological $N$-body simulation was performed to study the nonlinear perturbation evolution, see \cite{ref:HeFRnonliear,ref:MGhalofit} for examples, where the range of $f_{R0}$ was limited in $|f_{R0}| \lesssim10^{-4}$. However checking the linear matter power spectrum in this specific family of $f(R)$ models, as shown in Figure \ref{fig:pks}, one will find out the sensitive dependence on the values of model parameter $f_{R0}$ (or $D$, $B_0$). For future using the {\bf MGHalofit} to fit the nonlinear matter power spectrum, in this paper, we will take $f_{R0}$ as a free model parameter, then $D$ and $B_0$ are derived model parameters via the Eq. (\ref{eq:b0}) and Eq. (\ref{mG}) respectively. In Figure \ref{fig:pks}, the dependence to the model parameter $f_{R0}$ was shown, where the curves from the top to the bottom but the last one are plotted for the values of $f_{R0}$: $-10^{-2}$, $-10^{-3}$, $-10^{-4}$, $-10^{-5}$, $-10^{-6}$. The last one is for $\Lambda$CDM model. Here the other relevant cosmological parameters are fixed to their mean values obtained in {\it Planck} 2013 \cite{ref:Planck2013CP}. With this observations, one can expect to obtain a tight constraint to the model parameter $f_{R0}$ when RSD data points are included. If the values of $f_{R0}$ can be confined to a range less than $10^{-4}$ through the linear perturbation, one can safely use the {\bf MGHalofit} to obtain the nonlinear matter power spectrum prepared for comparison to the weak gradational lensing probes. 


When a modified gravity theory is confronted by the observed RSD $f\sigma_8(z)$ data points, the situation becomes really complicated. To obtain unbiased results, one should understand clearly the measured $f\sigma_8(z)$ and the methodology to make it (or the underlying assumptions). Here we give a brief review on how to obtain $f\sigma_8(z)$ by taking the WiggleZ survey as an example, please see \cite{ref:fsigma82-Blake2011} for the details. At first, the two-dimensional galaxy power spectrum $P_{g}(k,\mu)$ in different redshift slices is estimated using the familiar FKP method \cite{ref:FKP} using the fast Fourier transformation, where in mapping the angle-redshift survey cone into a cuboid of coming coordinates, a fiducial flat $\Lambda$CDM cosmological model with matter density $\Omega_m=0.27$ was used. Secondly, extracting the interested information, the growth rate $f$, a linear bias $b$  (and a variable damping coefficient $\sigma_v$, if it is included), by fitting the convolved power spectrum model with the window function to be compared to the data. Here the theoretical modelling of the observed data is crucial to improve the statistical accuracy and avoid systematic biases. The theory is well understood in the linear clustering regime \cite{ref:Kaisereffect}, but both simulations and observations have shown that the linear theory is a poor approximation in the quasi-linear scales encoding a great deal of clustering information. Therefore modelling power spectrum at the quasi-linear and nonlinear regimes deserves much efforts. The Ref. \cite{ref:fsigma82-Blake2011} fitted 18 power spectrum models based on the empirical non-linear velocity models, perturbation theory approaches and the fitting formulae from $N$-body simulations in the $\Lambda$CDM model. At last, the final model-independent results of $f\sigma_8(z)$ using the Jennings et. al. model \cite{ref:Jennings2011MNRAS,ref:Jennings2011APJ} were quoted, as collected in Table \ref{tab:fsigma8data}. Here one should note that this $f\sigma_8(z)$ is extracted under the assumption of $\Lambda$CDM model not only in the map making of $P_g(k,\mu)$ but also in the $P_{\delta\theta}$ and $P_{\theta\theta}$ calibration as a function of $z$ in terms of $P_{\delta\delta}$. One should also need to keep in mind that the $\sigma_8(z)$ is the rms fluctuation at the redshift $z$ of the {\bf linear} matter density field in comoving $8h^{-1}$ Mpc spheres as emphasised in Ref. \cite{ref:fsigma82-Blake2011}. It implies that the observed $f\sigma_8(z)$ should be the product of the linear growth rate and $\sigma_8(z)$ of the linear matter density field. Even though the growth rate for HS model and GR are consistent for extremely large scales $k<0.06h/\text{Mpc}$ at $z=0$ \cite{ref:frRSD}, this $f\sigma_8(z)$ cannot be used directly for a modified gravity theory due to the underlying assumption of $\Lambda$CDM model, say in fitting formulae calibration to $N$-body simulations in Jennings et. al. model \cite{ref:Jennings2011MNRAS,ref:Jennings2011APJ}. To have an unbiased $f\sigma_8(z)$ data points, one should repeat the above whole procedure for a modified gravity. But unfortunately it is still unavailable both in $N$-body simulation and data extraction now. The Ref. \cite{ref:frRSD} found a large deviation in the ratios $\sqrt{P_{\theta\theta}/P_{\delta\delta}}$ and $P_{\delta\theta}/P_{\delta\delta}$, which are the ratios of the growth rate in linear theory, between HS model and GR for $0.03<k/(h/\text{Mpc})<0.5$ in high resolution $N$-body simulations without complicated small scale damping. And there would be some differences for different $f(R)$ models. 

\begin{center}
\begin{figure}[tbh]
\includegraphics[width=9.6cm]{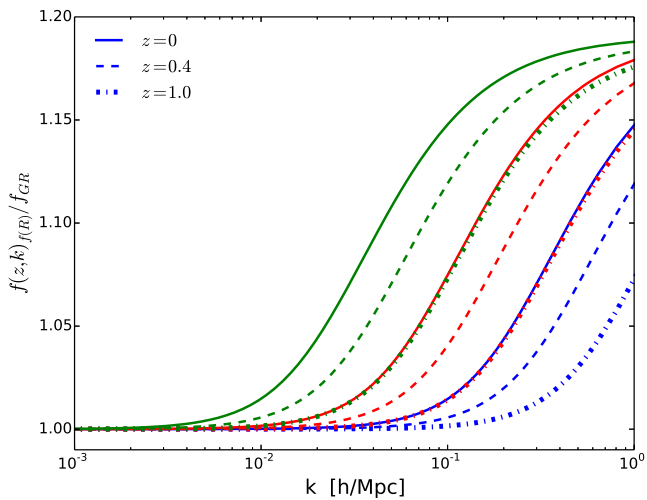}
\caption{The ratio of $f_{f(R)}/f_{GR}$ evolution with respect to $k$ at different redshift $z\in[0,1.0]$ for $f_{R0}=-10^{-6}$ (blue curves), $f_{R0}=-10^{-5}$ (red curves) and $f_{R0}=-10^{-4}$ (green curves) where the other relevant values are fixed to their mean values obtained in {\it Planck} 2013 \cite{ref:Planck2013CP}.}\label{fig:fk}
\end{figure}
\end{center}

Another thorny problem is the scale dependence of the growth rate $f(k,z)$ in a modified gravity, even in the range of linear scale which is determined by the model parameter $|f_{R0}|$. The scale dependence of the linear growth rate $f(z,k)$ for HS model was already shown in Ref. \cite{ref:frRSD} for different values of $|f_{R0}|$ at discrete redshift points. For this specific family of $f(R)$ models studied here, one can also see the similar scale dependence of $f(z,k)$ as shown in Figure \ref{fig:fk}, where large values of $|f_{R0}|$ predict large deviation to $\Lambda$CDM model. This confirms the results obtained in Ref. \cite{ref:frRSD}. But due to the degeneracy between the growth rate $f(z,k)$ and the galaxy bias $b$, it is not an easy task to find this sale dependence of $f(z,k)$. So when we compare the currently available $f\sigma_8(z)$ data points, which are sale independent and obtained based on $\Lambda$CDM cosmology, with the theoritical calculation, to alleviate the explicit scale dependence of $f\sigma_8(z)$, we should define $f\sigma_8(z)$ as  
\begin{equation}
f\sigma_8(z)=\frac{d\sigma_8(z)}{d\ln a},\label{ref:fsigma8def}
\end{equation}   
which is equivalent to the product of $f(z,k)$ and $\sigma_8(z)$ for a standard $\Lambda$CDM model. Because in a standard $\Lambda$CDM model case, the scale independence of linear growth rate $f(z)=d\ln D/d \ln a$ is due to the decomposition $\delta(z,k)\sim D(z)\delta(z=0,k)$. But it would not be the case for a modified gravity theory as studied in this paper. Hence using this compact and scale independent definition, i.e. the Eq. (\ref{ref:fsigma8def}), instead of the product of $f(z,k)$ and $\sigma_8(z)$ would be better and more universal. We expect it will be measured with this definition in the future.    
 
So what can one do with the $f\sigma_8(z)$ data on hand? With the observations on the Figure 11 in Ref. \cite{ref:frRSD}, say in the regime $k<0.1 h/\text{Mpc}$ at $z=0$ for $|f_{R0}|=10^{-4}$, the linear theory prediction for the growth rate almost matches the $N$-body simulation results for the $f(R)$ model, but deviates to the GR ones about $20\%$. Therefore, we naively assume that the underlying complication (including the scale dependence of the growth rate $f(z,k)$) can enlarge the error bars listed in Table \ref{tab:fsigma8data} to $20\%$, when the model parameter space is constrained.

\section{Results and Discussion} \label{sec:results}

In this section, we show the constraint results to the specific family of $f(R)$ models from the geometric and dynamic measurements. For the geometriccal one, we will use the supernova Ia data from SDSS-II/SNLS3 joint light-curve analysis \cite{ref:SNJLA}, the baryon acoustic oscillation $D_V(0.106) = 456\pm 27$ [Mpc] from 6dF Galaxy Redshift Survey \cite{ref:BAO6dF}; $D_V(0.35)/r_s = 8.88\pm 0.17$ from SDSS DR7 data \cite{ref:BAOsdssdr7}; $D_V(0.57)/r_s = 13.62\pm 0.22$ from BOSS DR9 data \cite{ref:sdssdr9}, the present Hubble parameter $H_0 = 73.8\pm 2.4$ [$\text{km s}^{-1} \text{Mpc}^{-1}$] from HST \cite{ref:HST}, and the full information of CMB recently released by {\it Planck}2013 (which include the high-l TT likelihood ({\it CAMSpec}) up to a maximum multipole number of $l_{max}=2500$ from $l=50$, the low-l TT likelihood ({\it lowl}) up to $l=49$) \cite{ref:Planckdata} with the addition of the low-l TE, EE, BB likelihood up to $l=32$ from WMAP9. For the dynamiccal one, we use the RSD data which was already listed in Table \ref{tab:fsigma8data}. For using the growth rate, we calculate the $f\sigma_8(z)=d\sigma_8/d\ln a$ at different redshifts in theory. 

\begin{center}
\begin{figure}[tbh]
\includegraphics[width=8.25cm]{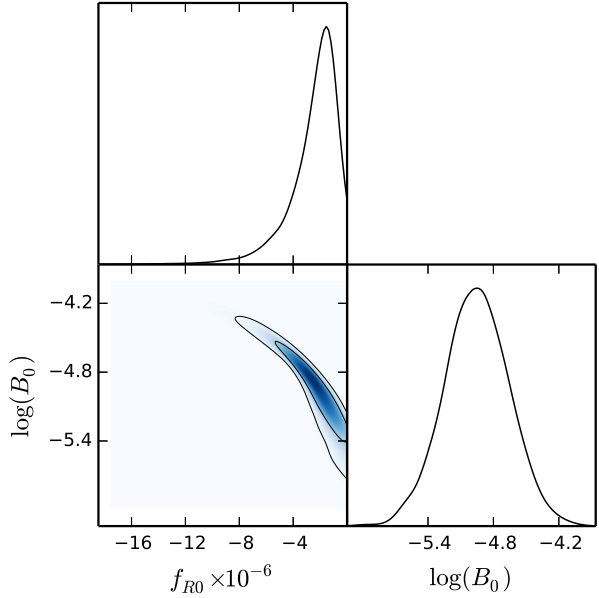}
\caption{The 1D marginalized distribution and 2D contours for interested model parameters with $68\%$ C.L., $95\%$ C.L. by using the {\it Planck} 2013, WMAP9, BAO, BAO, JLA, HST and RSD data sets.}\label{fig:contour}
\end{figure}
\end{center}

\begin{center}
\begin{figure}[tbh]
\includegraphics[width=9.5cm]{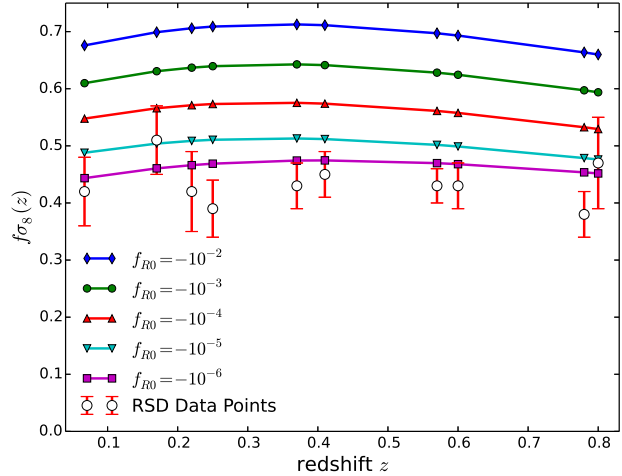}
\caption{The effects to $f\sigma_8(z)=d\sigma_8(z)/d\ln a$ for different values of $f_{R0}=-10^{-2}, -10^{-3}, -10^{-4}, -10^{-5}, -10^{-6}$ from the top to the bottom, where the values of other relevant cosmological model parameters were fixed to their best fit values listed in Table \ref{tab:results}.}\label{fig:fsigma8}
\end{figure}
\end{center}

We perform a global fitting on the {\it Computing Cluster for Cosmos} by using the publicly available package {\bf CosmoMC} \cite{ref:MCMC} in the following model parameter space
\begin{equation}
P=\{\Omega_b h^2,\Omega_c h^2,  100\theta_{MC}, \tau, n_s, {\rm{ln}}(10^{10} A_s),f_{R0}\},
\end{equation}
their priors are shown in the second column of Table \ref{tab:results}. The running was stopped when the Gelman \& Rubin $R-1$ parameter $R-1 \sim 0.02$ was arrived; that guarantees the accurate confidence limits. The obtained results are shown in Table \ref{tab:results} and Figure \ref{fig:contour}. 
\begingroup                                                                                                                     
\squeezetable                                                                                                                   
\begin{center}                                                                                                                  
\begin{table}                                                                                                                   
\begin{tabular}{cccc}                                                                                                            
\hline\hline                                                                                                                    
Parameters & Priors & Mean with errors & Best fit \\ \hline
$\Omega_b h^2$ & $[0.005,0.1]$ & $0.02242_{-0.00025}^{+0.00026}$ & $0.02239$\\
$\Omega_c h^2$ & $[0.001,0.99]$ & $0.1165_{-0.0015}^{+0.0015}$ & $0.1163$\\
$100\theta_{MC}$ & $[0.5,10]$  & $1.04169_{-0.00056}^{+0.00055}$ & $1.04198$\\
$\tau$ & $[0.01,0.81]$ & $0.079_{-0.012}^{+0.011}$ & $0.078$\\
${\rm{ln}}(10^{10} A_s)$ & $[2.7,4]$ & $3.056_{-0.022}^{+0.022}$ & $3.054$\\
$n_s$ & $[0.9,1.1]$ & $0.9687_{-0.0057}^{+0.0056}$ & $0.9702$\\
$f_{R0}\times 10^{-6}$ & $[-100,0]$ & $-2.58_{-0.58}^{+2.14}$ & $-1.36$\\  
\hline
$H_0$ & $...$ & $68.92_{-0.71}^{+0.68}$ & $69.05$\\
$\Omega_\Lambda$ & $...$ & $0.7060_{-0.0085}^{+0.0083}$ & $0.7078$\\
$\Omega_m$ & $...$ & $0.2940_{-0.0083}^{+0.0085}$ & $0.2922$\\
$\sigma_8$ & $...$ & $0.822_{-0.011}^{+0.011}$ & $0.816$\\
$z_{\rm re}$ & $...$ & $9.91_{-1.01}^{+0.98}$ & $9.88$\\
${\rm{Age}}/{\rm{Gyr}}$ & $...$ & $13.754_{-0.037}^{+0.037}$ & $13.748$\\
$fR_D$ & $...$ & $-0.0000044_{-0.0000010}^{+0.0000035}$ & $-0.0000023$\\
$\log(B_0)$ & $...$ & $-4.95_{-0.28}^{+0.29}$ & $-5.13$\\
\hline\hline                                                                                                     
\end{tabular}                                                                                                                   
\caption{The mean and best fit values with $1\sigma$ errors for the interested and derived cosmological parameters, where the {\it Planck} 2013, WMAP9, BAO, SN, HST and RSD data sets were used.}\label{tab:results}                                                                                                   
\end{table}                                                                                                                     
\end{center}                                                                                                                    
\endgroup 
The inclusion of RSD data set leads to very tight constraints on the model parameter $f_{R0}=-2.58_{-0.58}^{+2.14}\times 10^{-6}$ at $68\%$ C.L. (see Table \ref{tab:results}). In Figure \ref{fig:fsigma8}, we show the sensitive dependence of $f\sigma_8(z)$ on the model parameter $f_{R0}$, where one can see that the larger values of $f_{R_0}$ predict the larger values of $f\sigma_8(z)$. Actually, it is already seen from the linear matter power spectrum as shown in Figure \ref{fig:pks}. This is one of the main finding of this work. When this model parameter is well constrained on the linear scale, much time can be saved in $N$-body simulation by specifying the values obtained from the linear matter power spectrum.  

\begin{center}
\begin{figure}[tbh]
\includegraphics[width=9.5cm]{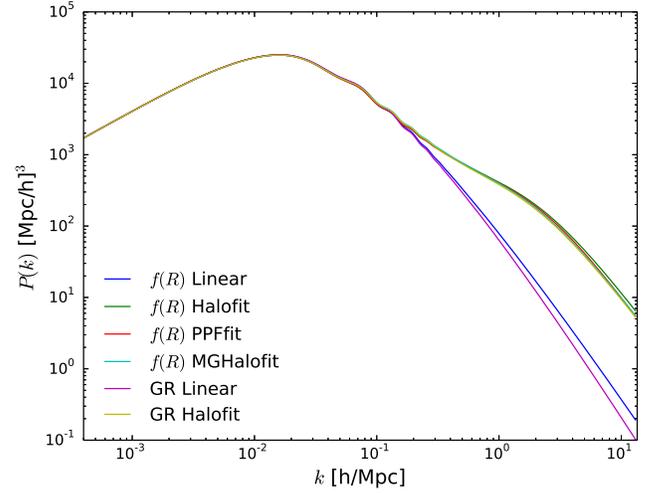}
\includegraphics[width=9.5cm]{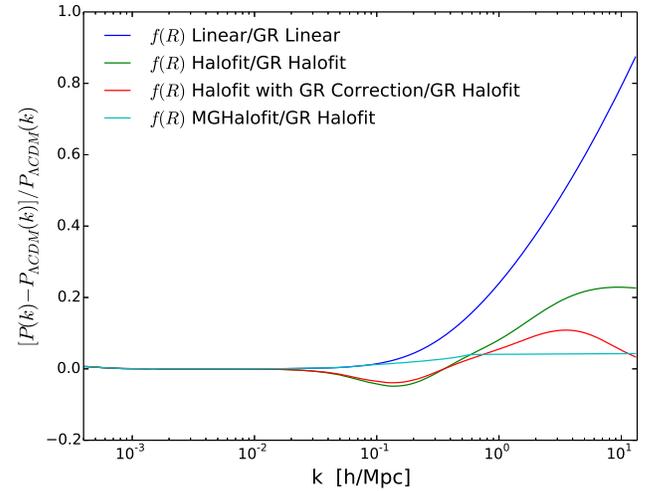}
\caption{The linear and nonlinear matter power spectrum at redshift $z=0$ for a specific family of $f(R)$ model with {\bf HALOFIT}, {\bf MGHalofit} and GR correction {\bf PPFfit} \cite{ref:HeFRnonliear} and that for $\Lambda$CDM model in GR, where the values of other relevant cosmological model parameters were fixed to their best fit values listed in Table \ref{tab:results}.}\label{fig:pksfrnl}
\end{figure}
\end{center}

Now let us move to the discussion of the nonlinear matter power spectrum at redshift $z=0$ obtained from {\bf HALOFIT}, {\bf MGHalofit} and the fitting formula given {\bf PPFfit} in Ref. \cite{ref:HeFRnonliear} where the other relevant cosmological model parameters were fixed to their best fitting values as obtained above in Table \ref{tab:results}. This comparison can provide clues to the difference of matter power spectrum at the nonlinear scales. We show the linear and nonlinear matter power spectrum corrected by different fitting in Figure \ref{fig:pksfrnl}. 

\begin{center}
\begin{figure}[tbh]
\includegraphics[width=9.5cm]{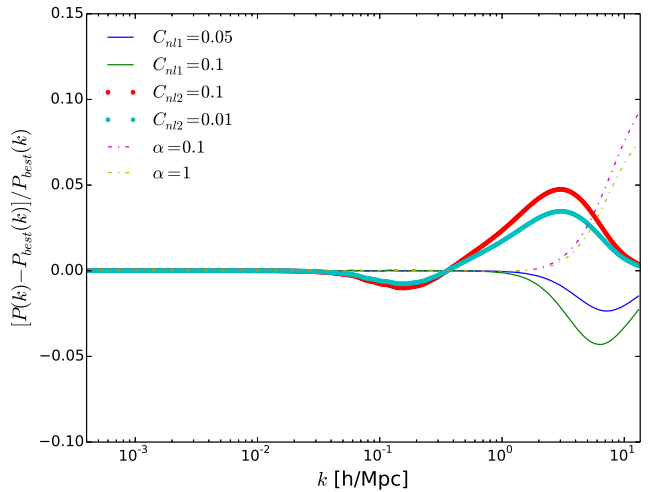}
\caption{The nonlinear matter power spectrum from {\bf PPFfit} at redshift $z=0$ for combinations of different values of $C_{\text{nl}1}$, $C_{\text{nl}2}$ and $\alpha$, where the values of other relevant cosmological model parameters were fixed to their best fit values listed in Table \ref{tab:results}.}\label{fig:pkshefitt}
\end{figure}
\end{center}

The top curve in Figure \ref{fig:pksfrnl} shows the relative difference of linear matter power spectrum between $f(R)$ and $\Lambda$CDM. This difference is due to the modification of the Newtonian constant $G$, that has been understood very well in Ref. \cite{ref:HeFRnonliear}. The second curve in Figure \ref{fig:pksfrnl} on the right side from the top was obtained from the standard {\bf Halofit} formula. That predicts relative large nonlinear matter power spectrum. The third curve in Figure \ref{fig:pksfrnl} on the right side from the top shows the relative difference of the nonleianr matter power spectrum corrected by GR nonlinear power spectrum from standard {\bf Halofit} model via the {\bf PPFfit} \cite{ref:HeFRnonliear}
\begin{equation}
P(k,z)=\frac{P_{{\rm non-GR}}(k,z)+\left(C_{{\rm nl}1}k^{\alpha}+C_{{\rm nl}2}\right)\Sigma^2(k,z)P_{{\rm GR}}(k,z)}{1+\left(C_{{\rm nl}1}k^{\alpha}+C_{{\rm nl}2}\right)\Sigma^2(k,z)},
\end{equation} 
where $P_{\rm GR}$ is the power spectrum in $\Lambda$CDM model and $\Sigma^2(k,z)$ is given by
\begin{equation}
\Sigma^2(k,z)=\left[\frac{k^3}{2\pi^2}P_{\rm lin}(k,z)\right]^{1/3},
\end{equation}
where $P_{\rm lin}(k,z)$ is the linear matter power spectrum in $f(R)$ gravity and $P_{{\rm non-GR}}(k,z)$ is the nonlinear power spectrum in $f(R)$ gravity without GR correction. Actually it is unknown priorly and the final nonlinear power spectrum what we are pursuing. Therefore, the linear matter power spectrum in a $f(R)$ gravity corrected by the standard {\bf HALOFIT} model is taken as a substitute. One can also find this kind of correction in Ref. \cite{ref:Hupk,ref:Koyamapk}. Here the values of $C_{\text{nl}1}=0.02349462$, $C_{\text{nl}2}=0.4634951$ and $\alpha=2.251794$ were adopted for the case of $f_{R0}=-10^{-4}$ at the redshift $z=0$, because the values of $C_{\text{nl}1}$, $C_{\text{nl}2}$ and $\alpha$ for the case of $f_{R0}=-10^{-6}$ at the redshift $z=0$ are still unavailable now. Therefore, we should keep in mind that different values of values of $C_{{\rm nl}1}$, $C_{{\rm nl}2}$ and $\alpha$ will change the shape and amplitude of the matter power spectrum at the nonlinear scale. To understand the changes, we plotted the nonlinear matter power spectrum with combination of different values of $C_{{\rm nl}1}$, $C_{{\rm nl}2}$ and $\alpha$ in Figure \ref{fig:pkshefitt}. The larger values of $C_{{\rm nl}1}$ will decrease the matter spectrum at the region larger than $k>1 h/{\rm Mpc}$. The larger values of $C_{{\rm nl}2}$ will increase the matter spectrum at the region larger than $k>0.2 h/{\rm Mpc}$. The larger values of $\alpha$ will decrease the matter spectrum at the region larger than $k>1 h/{\rm Mpc}$. Then choosing a combination carefully, a corrected power spectrum can mimic the evolution of the nonlinear matter power spectrum fitted from the {\bf MGHalofit} model or {\bf HALOFIT} model in $\Lambda$CDM model at a given redshift, say $z=0$. However, it is still hard to model the dependence of the model parameter $C_{{\rm nl}1}$, $C_{{\rm nl}2}$ and $\alpha$ to $f_{R0}$ at different redshifts \cite{ref:HeFRnonliear}. The {\bf MGHalofit} works in the range $|f_{R0}|\in[10^{-6},10^{-4}]$ and $z\le 1$, it is free from this kind of difficulties. As a comparison to the naive {\bf HALOFIT } and GR correction model, {\bf MGHalofit} predicts relative small deviation to the $\Lambda$CDM model based on {\bf HALOFIT}. One should worry about the suitability of {\bf MGHalofit} for this specific family of $f(R)$ models, because {\bf MGHalofit} is obtained based on Hu-Sawicki model \cite{ref:Hupk}, but for this tiny $|f_{R0}|\sim 10^{-6}$, it is difficult to detect a model not only because of the accuracy of the fitting formula but also because of the complicated astrophysical systematics on such scales \cite{ref:MGhalofit}. Based on these points, {\bf MGHalofit} would be a better choice.  

\section{Conclusion} \label{sec:conclusion} 

In this paper, a specific family of $f(R)$ models which can produce the $\Lambda$CDM background expansion history has been tightly constrained with an addition of the redshift space distortion data $f\sigma_8(z)$ combing the other cosmic observations which include SN, BAO, CMB and HST. Considering the scale dependence of the growth rate $f(z)$ for this this specific $f(R)$ model, we use the alternative definition $f\sigma_8(z)=d\sigma_8(z)/d\ln a$ in theory calculation, that is obviously scale independent and reproduces the conventional definition in the standard $\Lambda$CDM cosmology.  In doing so, under the assumption that future probes having the same best fit values as the current ten data points of $f\sigma_8(z)$, even having $20\%$ error bars enlarged, we find a preliminary constraint $f_{R0}=-2.58_{-0.58}^{+2.14}\times 10^{-6}$ in $1\sigma$ regions. This indicates the great potential that redshift space distortions have in constraining modified gravity theories. 

We have analyzed the nonlinear matter power spectrum at redshift $z=0$ in the specific family of $f(R)$ models using three fitting methods for the best fit values of model parameters. The first one is the standard {\bf HALOFIT} model, where the matter power spectrum deviates from the $\Lambda$CDM about $20\%$ at the nonlinear scales. The second is the {\bf PPFfit} method which is based on the {\bf HALOFIT} model with a correction from the $\Lambda$CDM model nonlinear power spectrum. At a fixed redshift say $z=0$, in principle, by carefully choosing values of model parameters $C_{{\rm nl}1}$, $C_{{\rm nl}2}$ and $\alpha$, almost the same nonlinear matter power spectrum as of $\Lambda$CDM model can be produced, but it is hard to model the dependence of $C_{{\rm nl}1}$, $C_{{\rm nl}2}$ and $\alpha$  on $f_{R0}$ at different redshifts. The third one is {\bf MGHalofit} model, although it is modeled based on Hu-Sawicki model and works in the range of $|f_{R0}| \in [10^{-6},10^{-4}]$ and $z\le 1$, the resultant nonlinear matter power spectrum can almost mimic the $\Lambda$CDM model with very small deviation from $\Lambda$CDM model. Also the dependence to the parameter $f_{R0}$ is well modeled. Based on this point, {\bf MGHalofit} would be a better choice, although it is based on the analysis of Hu-Sawicki model. With the very small values of $f_{R0}$, it is difficult to detect a model not only because of the accuracy of the fitting formula but also the complicated astrophysical systematics on such scales \cite{ref:MGhalofit}.

\acknowledgements{The author thanks an anonymous referee for helpful improvement of this paper and thanks Dr. Bin Hu for useful discussion and ICTP for hospitality during the author's visit in ICTP. This work is supported in part by National Natural Science Foundation of China under Grant No. 11275035 and No. 11491240169 (People's Republic of China), the Fundamental Research Funds for the Central Universities under Grant No. DUT13LK01, and the Open Project Program of State Key Laboratory of Theoretical Physics, Institute of Theoretical Physics, Chinese Academy of Sciences No. Y4KF101CJ1 (People's Republic of China).}

\end{document}